# Label-free electrical quantification of the dielectrophoretic response of DNA

Anja Henning, Jörg Henkel, Frank F Bier and Ralph Hölzel*

Address: Fraunhofer Institute for Biomedical Engineering, Am Mühlenberg 13, 14476 Potsdam-Golm, Germany

Email: Anja Henning - anjahenning@hotmail.com; Jörg Henkel - joerg.henkel@ibmt.fraunhofer.de; Frank F Bier - frank.bier@ibmt.fraunhofer.de; Ralph Hölzel* - ralph.hoelzel@ibmt.fraunhofer.de

* Corresponding author

## Abstract

A purely electrical sensing scheme is presented that determines the concentration of macromolecules in solution by measuring the capacitance between planar microelectrodes. Concentrations of DNA in the ng/mL range have been used in samples of 1 μL volume. The method has been applied to the characterisation of the dielectrophoretic response of DNA without the need for any chemical modifications. The influence of electrical parameters like duty cycle, voltage and frequency has been investigated. The results are in good agreement with data from dielectrophoretic studies on fluorescently labelled DNA. Extension of the method down to the single molecule level appears feasible.

PACS: 87.50.ch, 87.80.Fe, 87.85.fK

## Introduction

For the construction of systems on the nanometre scale there is a growing need for alternatives to classical photolithography. A promising approach for this is the exploitation of the self-organising properties of biological macromolecules, in particular DNA (deoxyribonucleic acid) [1,2]. Double-stranded DNA consists of two DNA single strands which form the double helix. It is stabilised by hydrogen bonds between complementary purine and pyrimidine bases. This coupling is very specific and allows to address distinct sites on the DNA at a resolution of 0.34 nm (for B-DNA), i.e. the distance between neighbouring bases. Addressing can easily be accomplished by chemical means in arbitrary volumes, hence in an extremely parallelised manner. Most methods for the synthesis and modification of DNA are well established in molecular biology. However, the characterisation of these constructs and their connection to the macroscopical world are still demanding. For optical detection fluorescent markers are common. Still, labelling of the mole-cules is necessary, and bleaching of the fluorophores leads to artefacts and limits the possible observation time. Electrochemical sensing calls for chemical modifications, too, either of the target molecules or of the electrodes [3-5]. Label-free characterisation on surfaces is possible e.g. by scanning probe microscopy [6,7] and optical methods like surface plasmon resonance and grating couplers [8,9]. Highly desirable would be a purely electrical detection scheme. This is because such a principle could be well integrated into lab-on-a-chip systems, neither optical nor mechanical access would be necessary, and geometrical resolution would principally not be restricted as it is the case with optical methods. Here we present a purely electrical sensing scheme based on the measurement of capacitance changes between microelectrodes caused by DNA concentration changes.

These variations in local DNA concentration are also achieved by electrical means applying dielectrophoresis (DEP). Here an inhomogeneous electrical AC field exerts forces onto macromolecules like DNA towards the electrode edges [10-12]. This method is increasingly exploited for the concentration and alignment of nano-objects like DNA, proteins, nano wires and carbon nanotubes [13,14]. Whilst it is widely applied as a micro- and nano-tool [15-17] there are only few studies aimed at a fundamental understanding of molecular DEP [18-21]. Therefore we have used the presented sensing scheme for quantifying the dielectrophoretic response of DNA. In contrast to all other known studies on molecular dielectrophoresis of DNA there is no need for any fluorescent labelling of the sample.

**Methods**

The electrode chamber has been prepared from commercially available surface acoustic wave resonators (R2633, Siemens/Matsushita). Their characteristic frequency of 433.6 MHz lies far away from the frequencies chosen in this study. Therefore the influence of surface waves can be neglected here. A quartz substrate of 4 mm length, 1 mm width and 0.5 mm height carries two pairs of 300 nm thick interdigitated aluminium electrodes. Each electrode consists of 35 fingers of 800 μm length and 2.3 μm width leaving an interelectrode gap of 1.7 μm (Fig. 1) [22]. A silicon rubber gasket of 0.5 mm thickness was trimmed using a $CO_2$ laser plotter (Epilog Laser Legend 24TT) and mounted around the substrate with double-sided adhesive tape. It was sealed with a cover glass and immersion oil. Fluid samples of 12 μL volume were pipetted onto the electrodes leaving an air-filled space between fluid and gasket. Thus the sample only came into contact with the electrodes, the quartz substrate and the cover glass. This helped to minimise contaminations which can easily occur due to the sample's high surface-to-volume ratio.

Dielectrophoresis and impedance measurements have first been combined by Milner et al. [23] and Suehiro et al. [24] for the characterisation of bacteria. They applied a lock-in amplifier or an oscilloscope for the determination of phase and amplitude and, hence, impedance. Arnold [25] used an impedance analyser for studying the DEP behaviour of yeast cells. In the simplest case the DEP field signal also served as the source signal for the impedance measurement. Alternatively, two signal sources were used in order to choose the properties of both signals independ-ently. This made it possible to select the measuring frequency for optimal sensitivity. However, this led to the need for additional electronic circuitry in order to combine both signals.

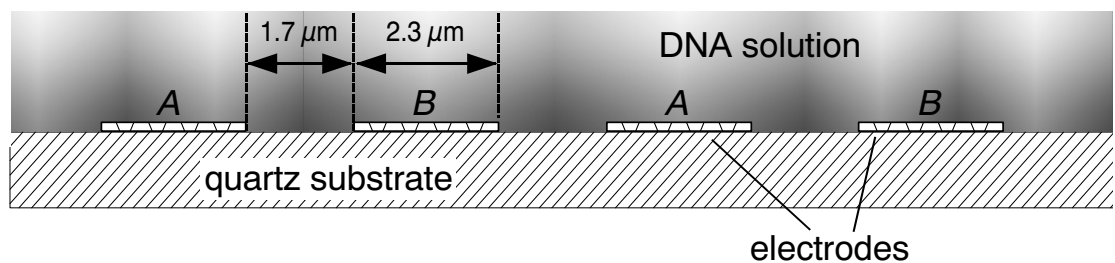

**Figure 1**
**Cross section of the interdigitated electrodes**. Each aluminium electrode (A, B) consists of 35 fingers of 800 μm length. The shading of the DNA solution is meant to illustrate the possible DNA attraction during dielectrophoresis.

Usually transformers were used limiting the accessible frequency range as well as the amplitude range. Recently Beck et al. [26] introduced an operational amplifier for this purpose with, however, similar restrictions. Subsequently both signals had to be separated for impedance determination calling for additional low-pass filters [23,26] or a properly balanced bridge [25]. Hölzel and Bier [27] introduced a somewhat different approach by separating DEP and measurement temporally by switching. They used a frequency variable RLC-meter rendering any additional circuitry unnecessary. However, sensitivity and DEP amplitude were limited. For a significant improvement, in particular of the sensitivity, in this work the measurement was performed with an ultra-precision capacitance bridge (Andeen-Hagerling AH 2550A). It was connected to the micro-electrodes through relays (Fig. 2) and was controlled by a personal computer using purpose-built software. The reed relays (Meder) also were computer controlled via the capacitance bridge. The measuring signal of 1 kHz was kept at or below 30 mV$_{RMS}$ to minimise impact on the measurement itself. The dielectrophoresis signal was supplied by an RF synthesizer (Hameg HM 8133-2) and raised to up to 17 V$_{RMS}$ by a power amplifier (Toellner Toe 7606). The synthesizer output could be modulated or gated by a DDS generator (TTi TG 1010 A) giving a variable duty cycle of the DEP field between 0.1% and 100%. DEP field amplitude and duty cycle were monitored by a digital oscilloscope (Agilent MSO 6104 A). All electrical connections were shielded. The input cables to the capacitance bridge were double shielded and arranged close to each other to minimise loop area and, hence, magnetically induced pickup.

As DNA sample the phagemid pBluescript was used. It has a length of 2961 base pairs corresponding to 1.0 μm contour length. That means that it did not bridge the electrode gap of 1.7 μm. pBluescript was prepared from a transformed *E. coli* culture and linearised by digestion with the restriction enzyme Eco RI. It was purified using an Invisorb Spin PCRapid kit (Invitek) and diluted with deionised water to final concentrations ranging from 18 pM to 18 nM.

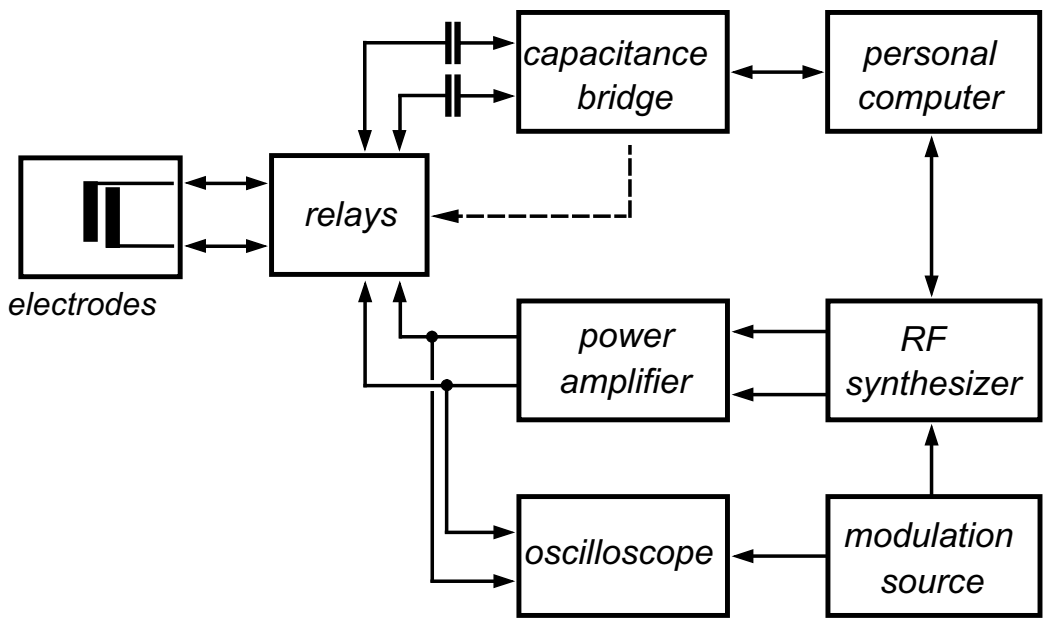

**Figure 2**
**Electrical setup for combined dielectrophoresis and impedance measurement**. Electrodes are alternately connected by relays either to the capacitance bridge for measurement or to the DEP signal supply for dielectrophoretic action. A personal computer controls relays, capacitance bridge and RF synthesizer.

## Results and discussion

For the quantification of molecular DEP response the dielectrophoresis field was applied to the electrodes for 8 s followed by a measuring period of 2 s. As a measure of DEP response the increase in capacitance during DEP was taken. In order to examine whether this is a suitable measure, DEP frequency and amplitude were kept constant at 1 MHz and 4 $V_{RMS}$, resp., and the field was 100% square modulated, that means it was switched on and off, at a rate of 1 kHz. The modulation's duty cycle was varied from 0.1% to 100% (Fig. 3). This should result in dielectrophoretic action onto the DNA being proportional to the duty cycle.

As can be seen from Fig. 4, the determined capacitance changes ΔC increased nearly linear with DEP action up to 1.6 pF at a duty cycle of 30%, being saturated at higher values. A similar saturation effect for pBluescript has been reported by Du et al. [18] using fluorescence changes as a measure of DEP response. In order to ensure a linear relation between DEP action and capacitance changes, in the following experiments conditions were chosen that resulted in ΔC values within this range of 1.6 pF.

The time constant of the decrease in capacitance due to diffusion of the DNA after DEP application was found to amount to 30 s at 18 nM DNA concentration (Fig. 3, t = 120 s...240 s). This

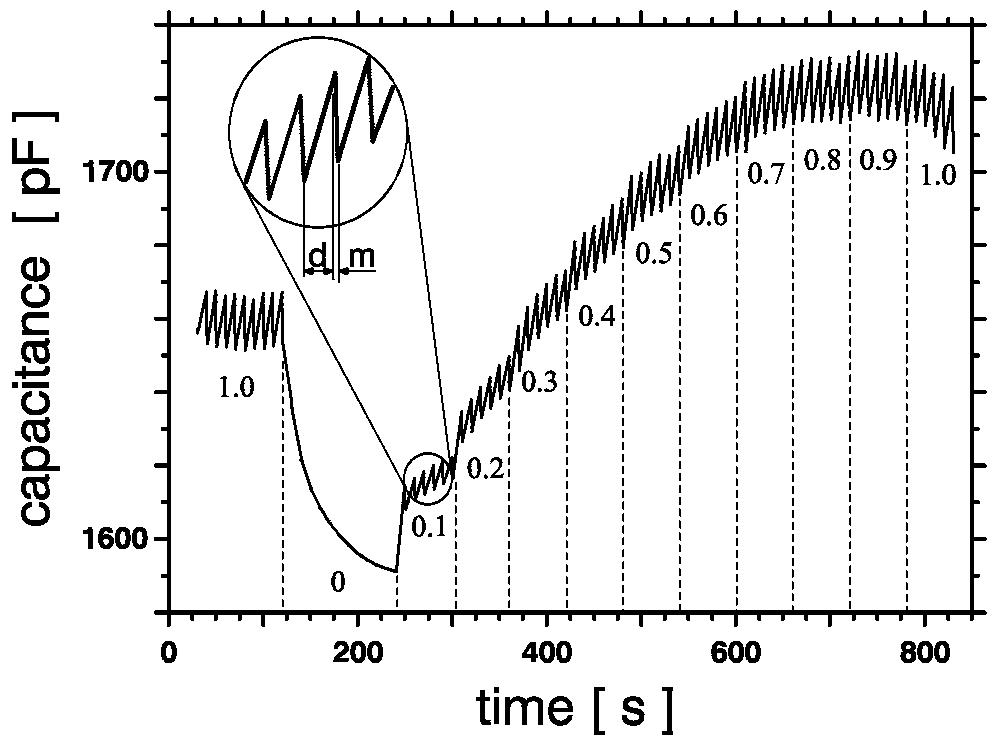

#### Figure 3
**Time course of capacitance for a typical DEP experiment**. Capacitance of an aqueous solution of 18 nM pBluescript DNA with a DEP excitation of 1 MHz and $4V_{RMS}$. The duty cycle of the modulated signal was varied from 0.1% ("0") to 100% ("1.0"). The magnified view shows the change between 8 s DEP ("d") application and 2 s measurement ("m").

decrease is much slower than the data acquisition speed which is 200 ms or less for each capacitance data. However, this variation in capacitance still limits the resolution of the presented method. This is because the actual capacitance changes take place on a shorter time scale than the bridge's automatic balancing procedure for complete balancing. The observed capacitance decay of 80 pF is an order of magnitude larger than the typical capacitance change recorded for DEP quantification. This means that by far most of the measured signal is a consequence of reversible DNA attraction and that most of the DNA does not adhere to the electrodes.

This is inconsistent with the observations of Washizu et al. [28] who report spontaneous permanent fixation of DNA to aluminium. On the other hand this result agrees well with the work of Kabata et al. [15] who deliberately functionalised the ends of DNA with avidin in order to achieve permanent adherence to the aluminium electrodes.

The values for capacitance change at 100% duty cycle increased in the course of the experiment by 9% ($t_1$ = 100 s, $t_2$ = 800 s). That means that there is a systematic error introduced, presumably by the gradual concentration increase close to the electrodes due to DEP action, which also is reflected by the increase in the absolute average capacitance by 4% within this period. Errors are also introduced by aggregation of DNA molecules [19,29] which leads to variations of the local DNA concentration in the electrode gap [30]. As a consequence of the DEP duration being much shorter than the time needed to reach equilibrium concentration the determined capacitance changes reflect the initial dielectrophoretic collection rate as described by Bakewell and Morgan [19], whilst in other studies steady-state DNA concentrations were monitored [18,31].

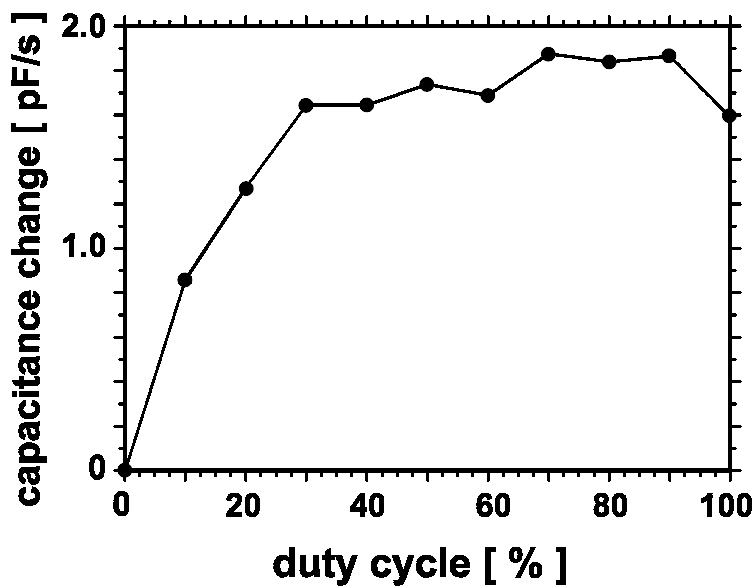

**Figure 4**
**Capacitance change as a function of DEP field duty cycle**. Data of Fig. 3 are presented as a function of the duty cycle of the dielectrophoresis field. pBluescript DNA concentration was 18 nM. Dielectrophoretic excitation was at a frequency of 1 MHz with a voltage of $4V_{RMS}$.

In the typical combination of impedance measurement and dielectrophoresis biological cells are usually studied, which have to be agitated after or during each measurement to achieve an even distribution for the following measurement. For this end flow through systems are used [23,24]. When extending the method onto molecules mixing already occurs by thermally driven stochastic fluctuations (Brownian motion). Consequently a batch system is also suitable. Such a static system is not only simpler, it also requires much smaller sample volumes. Therefore it is possible to reduce the current volume of about 1 μL even further.

Zheng et al. [21] investigated the suitability of DNA and proteins for the manufacturing of electronic devices. For this purpose they measured the electrical resistance between narrow electrode tips during and after dielectrophoretic manipulation of DNA and of the protein BSA, however, without any success. Probably this was mainly a consequence of the small interaction length of their tip electrodes of only about 10 μm as compared to the interdigitated electrode arrange-ment of this work, which extends over 55 mm with only a fifth of the gap width. Above this, we observed that concentration changes of DNA led to a more distinct response in the capacitive part of the impedance than in its resistive part.

The electrical conductivity of the sample solution strongly influences the dielectrophoretic response. Due to the small volumes involved and, hence, high surface-to-volume ratio it is problematic to presume equal conductivities in the stock solution and in the actual sample *in situ*. Even the stock solution itself is too small to be measured with a standard conductivity probe. We therefore used the resistive part of the bridge's output to deduce the electrical conductivity of the actual sample volume after calibration with solutions of known conductivity. From this an upper bound of 5 mS/m for the electrical conductivity in all experiments follows. From AC electrokinetic studies on similarly small sample volumes [32,33] a lower bound of 1 mS/m can be deduced.

The stability of the DNA double helix is influenced by the solution's salt content. At low salt concentrations double stranded DNA tends to disintegrate into its composing single strands. This is reflected by the melting point of hybridised DNA, which can be approximately calculated from its base composition and the salt concentration [34]. The latter can be estimated from the sample's electrical conductivity to lie in the range between 0.08 mM and 0.4 mM. This results in a melting point of pBluescript between 34°C and 45°C. The melting point is also influenced by the solution's pH-value. For deionised water it is around pH 5 due to atmospheric $CO_2$. This will lower the melting point by about 2°C [35] to 32°C to 43°C. This is well above the experimental temperature of 22 (± 2) °C and in accordance with other dielectrophoretic studies on double stranded DNA having been performed in deionised water [21,36]. Still, it cannot fully ruled out for this work as well as for other dielectrophoretic studies that a minor portion of DNA is present as single strands.

In order to quantify the dependence of molecular dielectrophoresis on field strength DEP voltage was varied from 2 $V_{RMS}$ to 4.5 $V_{RMS}$ at a fixed frequency of 1 MHz (Fig. 5). From the slope of the double logarithmic plot clearly a cubic relation follows, which is in contradiction to generally accepted DEP theory [37,38]. DEP action usually is explained by dipoles that are induced by the DEP field and interact with this field, resulting in a square dependence on field strength. Similar experiments on fluorescently labelled pBluescript DNA also showed a cubic relation at 1 MHz [18], whilst the data of Asbury et al. [39] at 30 Hz are more consistent with a square dependence. Tuukkanen et al. [31] have investigated the trapping efficiency for DNA samples of various lengths as a function of applied DEP voltage. They found a deviation from a purely quadratic relation and interpreted this by assuming a threshold resulting from thermal drag force with which the dielectrophoretic force competes. They also mentioned the well known fact that DNA longer than a few hundreds of basepairs is in a globular shape as long as there are no external forces present, and that DEP leads to elongation of the molecules and, hence, to an enhanced polaris-ability. In our view this mechanism of enhanced polarisability alone is sufficient to account for the observed more-than-quadratic DEP-voltage dependence. It means that in the equation for the dielectrophoretic force $F_{DEP}$ the polarisability $\alpha$ is no longer a constant but a function of the electric field $E$ itself: $F_{DEP} = \alpha(E) \cdot á\,(E^2)$. Bakewell and Morgan [19] reported DEP collection data from fluorescently labelled supercoiled plasmid DNA that deviated from a purely square relation. They, too, considered a change in plasmid shape under the action of a high DEP force and, additionally, discussed the action of fluid flow that is caused by electrohydrodynamic (EHD) forces, in particular AC-electroosmosis [40,41]. When re-evaluating the data of Tuukkanen et al. [31] by plotting their voltage-fluorescence relation in a double-logarithmic plot (data not shown) we got slopes between 2.9 and 3.8 ($r^2$ = 0.723...0.994) for DNA lengths ranging from 27 bp to 8461 bp. Therefore the origin of the deviation from a purely quadratic dependence between electric field and molecular DEP response remains ambiguous. For a further clarification the DEP response of globular molecules like proteins should be quantified or, even better, stable compact DNA constructs like origami structures [42,43] should be studied. Variation of the DEP field's duty cycle especially in the region of small capacitance changes would surely clarify the influence of threshold effects.

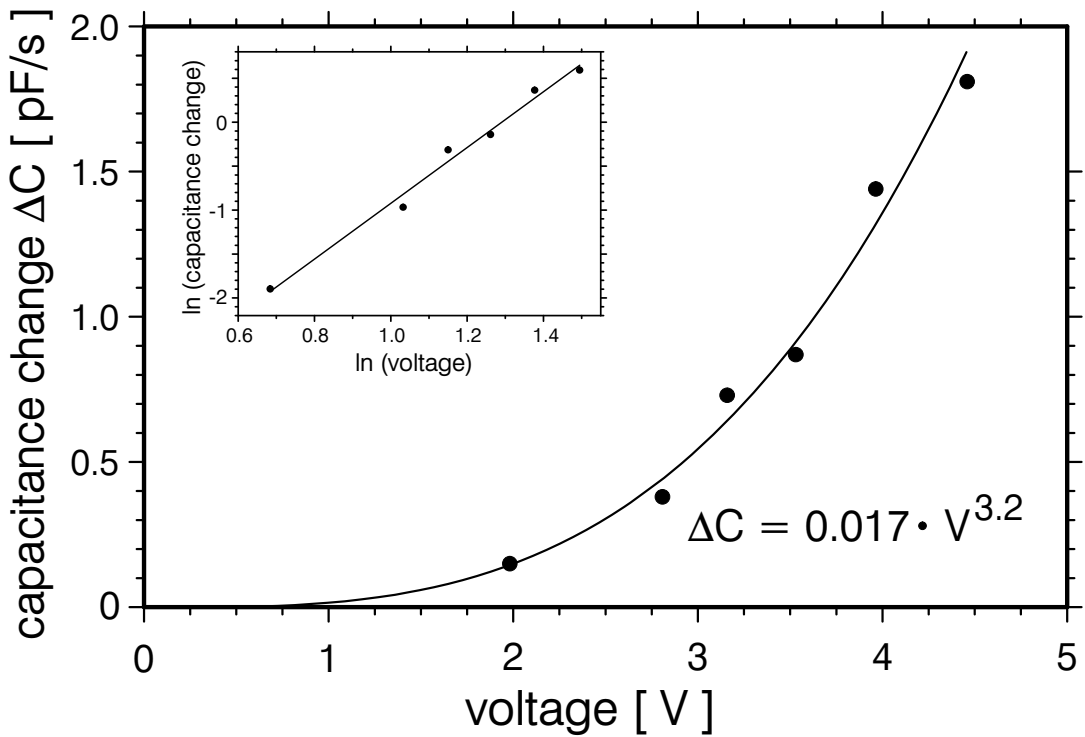

**Figure 5**
**Capacitance changes as a function of DEP voltage**. Dielectrophoretic excitation was performed at a frequency of 1 MHz. pBluescript DNA concentration was 18 nM. The inset shows the data in a double logarithmic plot.

For a better understanding of molecular dielectrophoresis we have measured its frequency dependence between 1 kHz and 3 MHz (Fig. 6). The electrically determined DEP response agrees very well with the fluorescence data of Tuukkanen et al. [31] who found positive dielectrophoresis at a rather constant level between 100 kHz and 10 MHz. The findings of Bakewell and Morgan [19] and of Du et al. [18] are slightly different showing a stronger decrease of DEP with increasing frequency. However, differences in the experimental setup, e.g. in the conductivity of the DNA solution and in parasitic inductances, might account for these deviations. At frequencies below 30 kHz the measured capacitance changes became negative indicating negative dielectrophoresis. This is contradictory to the results of Chou et al. [44] who observed positive DEP below 1 kHz. However, they used an electrodeless setup with more confined attraction regions thereby avoiding disturbing effects like electrolysis and fluid flow. Fluid flow due to nonuniform AC fields has been investigated in detail by Green et al. [40,41] for an electrode geometry similar to the interdigitated structure of this work. They found strong flow by AC electroosmosis at field frequencies between about 100 Hz and 10 kHz coinciding well with the range of apparent negative capacitance change found here. For a better understanding variations of the field's duty cycle in this frequency range would be helpful as well as simultaneous microscopical observation of fluorescently labelled DNA.

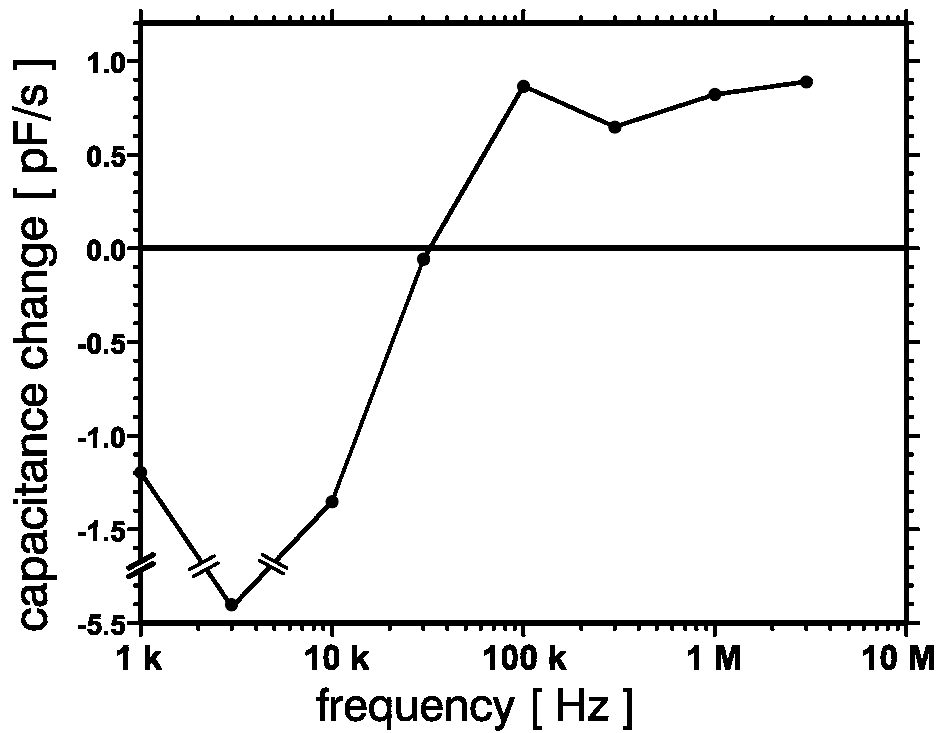

**Figure 6**
**Dielectrophoretic spectrum of pBluescript DNA**. DNA concentration was 18 nM. DEP amplitude was 5.6 $V_{RMS}$.

The suitability of the present setup for a quantification of macromolecular concentrations in general is shown in Fig. 7. The concentration of pBluescript DNA was varied between 1.8 pM and 1.8 nM, corresponding to 3.5 ng/ml and 3.5 μg/ml, resp. The semi-logarithmic plot shows a good correlation between DNA concentration and capacitance. A similar result has been reported by Macanovic et al. [45] using chemically modified electrodes of 1 cm$^2$ surface area, however, at DNA concentrations that were at least a factor of 30 higher (0.1 μg/ml to 2 μg/ml). The dielectrophoretic response of DNA strongly depends on its molecular weight with higher field strengths being needed for smaller molecules [31]. Still, DEP can be applied successfully to DNA of contour lengths well below a tenth of the electrode spacing as well as to DNA spanning this gap more than 20 times. The size dependence could be exploited for the sensing of molecular weight by applying more complex electrode structures and excitation schemes.

It is of interest to estimate whether the present setup should be capable of detecting single macromolecules. The capacitance bridge used in this work is specified by the manufacturer with a reportable resolution of $10^{-7}$ pF under optimal conditions. If one considers a maximal measuring voltage of 30 mV$_{RMS}$ as used in this work the resolution should be reduced to about $10^{-4}$ pF. This is three orders of magnitude better than has been achieved in this work and is mainly a con-sequence of the fast concentration changes after dielectrophoretic attraction. Additionally, the shielding of the cables currently used is not better than 60 dB and the contact-to-contact capacitance of the relays amounts to about 1 pF. Both factors allow interference by external noise sources. From the present data it follows, that under optimal conditions and if one considers electrodes of e.g. 1 µm width and a mutual distance of 1 µm a single pBluescript molecule would result in a capacitance change of more than $10^{-2}$ pF, which is still two orders of magnitude above the calculated resolution. Therefore it appears quite feasible to further develop the presented apparatus towards an electronic labelfree single molecule detector. Selectivity would be readily achieved by chemical modifications of the electrodes or of the gap by e.g. complementary DNA or antibodies as it is common practise in other labelfree detection schemes like surface plasmon resonance and grating couplers [8,9]. Additionally, such a detector could be used in combination with a dielectrophoretic single molecule trap [46] allowing for an automatic consecutive investigation of large numbers of single molecules.

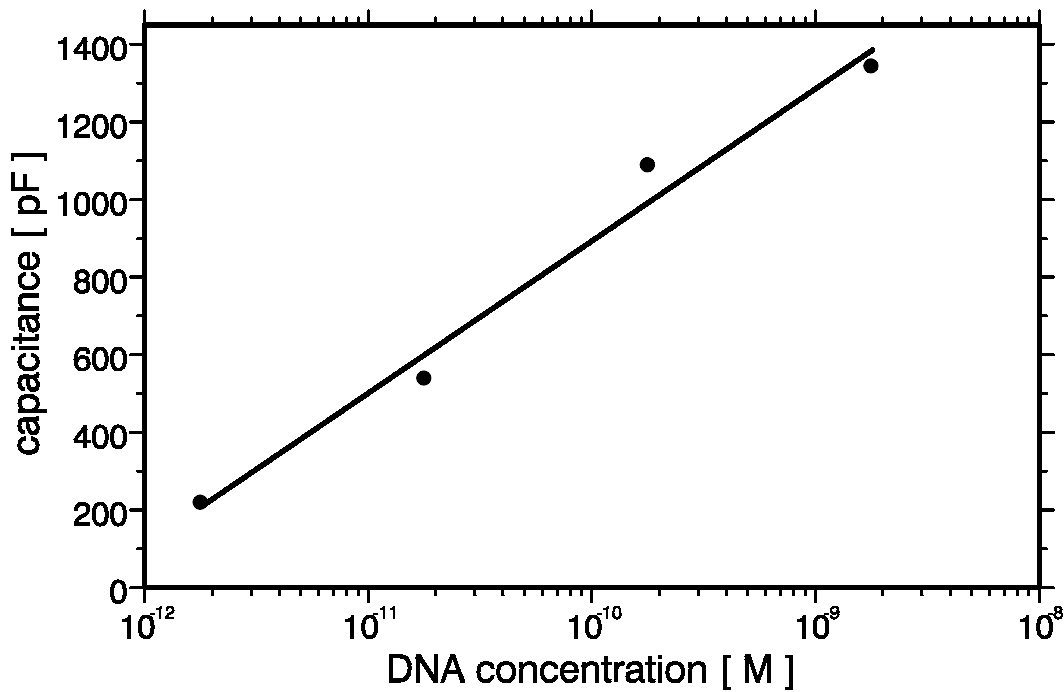

**Figure 7**
**Capacitance as a function of pBluescript DNA concentration**. Capacitance of deionised water alone was 15 pF.

## Conclusion

A system has been developed for the measurement of the concentration of macromolecules by monitoring the capacitance between interdigitated electrodes. It has been applied to the determination of the dielectrophoretic response of DNA without the need for any chemical modification of the analyte. Being purely electronic the method can be easily integrated into lab-on-a-chip systems. Neither optical nor mechanical access to the sample is needed in the course of the measurement. An improvement of the temporal resolution by about an order of magnitude appears rather straighforward. Some modifications of the experimental design will allow for a downscaling of the actual sample volume to a few µm$^3$ or even less. In this case the resolution of the instrumentation will be adequate to automatically detect and possibly characterise single macromolecules.

### Acknowledgements
We would like to thank Andreas Calender for assistance in software design and Mandy Lorenz for her help concerning DNA preparations. Financial support by the European Communities within the project Nucan (STRP 013775) is gratefully acknowledged.

### References
1. Carbone A, Seeman NC: *Proc Nat Acad Sci (USA)* 2002, **99:**12577-12582.
2. Yan H, Park SH, Finkelstein G, Reif JH, LaBean TH: *Science* 2003, **301:**1882-1884.
3. Drummond TG, Hill MG, Barton J: *Nature Biotechnol* 2003, **21:**1192-1199.
4. Park S-J, Taton TA, Mirkin CA: *Science* 2002, **295:**1503-1506.
5. Xiao Y, Lubin AA, Baker BR, Plaxco KW, Heeger AJ: *Proc Nat Acad Sci (USA)* 2006, **103:**16677-16680.
6. Hörber JKH, Miles MJ: *Science* 2003, **302:**1002-1005.
7. Kada G, Kienberger F, Hinterdorfer P: *Nano Today* 2008, **3:**12-19.
8. Homola J: *Anal Bioanal Chem* 2003, **377:**528-539.
9. Gauglitz G: *Anal Bioanal Chem* 2005, **381:**141-155.
10. Washizu M, Kurosawa O: *IEEE Trans Ind Appl* 1990, **26:**1165-1172.

11. Hölzel R, Bier FF: *IEE Proc Nanobiotechnol* 2003, **150:**47-53.
12. Morgan H, Green NG: *AC electrokinetics: colloids and nanoparticles* Baldock, Research Studies Press; 2003.
13. Lapizco-Encinas BH, Rito-Palomares M: *Electrophoresis* 2007, **28:**4521-4538.
14. Stokes P, Khondaker SI: *Nanotechnology* 2008, **19:**175202.
15. Kabata H, Kurosawa O, Arai I, Washizu M, Margarson SA, Glass RE, Shimamoto N: *Science* 1993, **262:**1561-1563.
16. Krupke R, Hennrich F, v Löhneysen H, Kappes MM: *Science* 2003, **301:**344-347.
17. Liu X, Spencer JL, Kaiser AB, Arnold WM: *Curr Appl Phys* 2006, **6:**427-431.
18. Du M-L, Bier FF, Hölzel R: *AIP Conf Proc* 2006, **859:**65-72.
19. Bakewell DJ, Morgan H: *IEEE Trans Nanobioscience* 2006, **5:**1-8.
20. Salonen E, Terama E, Vattulainen I, Karttunen M: *Eur Phys J E* 2005, **18:**133-142.
21. Zheng L, Brody JP, Burke PJ: *Biosens Bioelectron* 2004, **20:**606-619.
22. Bier FF, Gajovic-Eichelmann N, Hölzel R: *AIP Conf Proc* 2002, **640:**51-59.
23. Milner KR, Brown AP, Allsopp DWE, Betts WB: *Electronics Lett* 1998, **34:**66-68.
24. Suehiro J, Yatsunami R, Hamada R, Hara M: *J Phys D: Appl Phys* 1999, **32:**2814-2820.
25. Arnold WM: *IEEE Conf Proc Electrical Insul Diel Phenom* **2001:**40-43.
26. Beck JD, Shang L, Marcus MS, Hamers RJ: *Nanolett* 2005, **5:**777-781.
27. Hölzel R, Bier FF: *AIP Conf Proc* 2004, **725:**77-83.
28. Washizu M, Kurosawa O, Arai I, Suzuki S, Shimamoto N: *IEEE Trans Ind Appl* 1995, **231:**447-456.
29. Washizu M, Suzuki S, Kurosawa O, Nishizaka T, Shinohara T: *IEEE Trans Ind Appl* 1994, **30:**835-843.
30. Hölzel R: *J Electrostat* 2002, **56:**435-447.
31. Tuukkanen S, Kuzyk A, Toppari JJ, Häkkinen H, Hytönen VP, Niskanen E, Rinkiö M, Törmä P: *Nanotechnol* 2007, **18:**295204.
32. Hölzel R: *Biochim Biophys Acta* 1999, **1450:**53-60.
33. Hölzel R: *Biophys J* 1997, **73:**1103-1109.
34. Sambrook J, Russell DW: *Molecular cloning: A laboratory manual* New York, Cold Spring Harbor Laboratory Press; 2001.
35. Privalov PL, Ptitsyn OB, Birshtein TM: *Biopolymers* 1969, **8:**559-571.
36. Asbury CL, Engh G van den: *Biophys J* 1998, **74:**1024-1030.
37. Pohl HA: *Dielectrophoresis* Cambridge, Cambridge University Press; 1978.
38. Wang X-B, Huang Y, Hölzel R, Burt JPH, Pethig R: *J Phys D: Appl Phys* 1993, **26:**312-322.
39. Asbury CL, Diercks AH, Engh G van den: *Electrophoresis* 2002, **23:**2658-2666.
40. Green NG, Ramos A, Gonzalez A, Morgan H, Castellanos A: *Phys Rev E* 2000, **61:**4011-4018.
41. Green NG, Ramos A, Gonzalez A, Morgan H, Castellanos A: *Phys Rev E* 2002, **66:**026305.
42. Rothemund PWK: *Nature* 2006, **440:**297-302.
43. Kuzyk A, Yurke B, Toppari JJ, Linko V, Törmä P: *Small* 2008, **4:**447-450.
44. Chou C-F, Tegenfeldt JO, Bakajin O, Chan SS, Cox EC, Darnton N, Duke T, Austin RH: *Biophys J* 2002, **83:**2170-2179.
45. Macanovic A, Marquette C, Polychronakos C, Lawrence MF: *Nucl Acid Res* 2004, **32:**e20.
46. Hölzel R, Calander N, Chiragwandi Z, Willander M, Bier FF: *Phys Rev Lett* 2005, **95:**128102.